\pdfoutput=1
\documentclass[prl,superscriptaddress,aps,preprint]{revtex4}

\usepackage{amsmath}
\usepackage{amssymb}
\usepackage{latexsym}
\usepackage{graphicx}
\usepackage{epsfig}
\usepackage{epstopdf}

\begin{document}
\title{Observation of bright polariton solitons in a semiconductor microcavity}

\author{M.~Sich}
\affiliation{Department of Physics and Astronomy, University of Sheffield, Sheffield S3 7RH, United Kingdom}

\author{D.~N.~Krizhanovskii}
\affiliation{Department of Physics and Astronomy, University of Sheffield, Sheffield S3 7RH, United Kingdom}

\author{M.~S.~Skolnick}
\affiliation{Department of Physics and Astronomy, University of Sheffield, Sheffield S3 7RH, United Kingdom}

\author{A.~V.~Gorbach}
\affiliation{Department of Physics, University of Bath, Bath, BA2 7AY, United Kingdom}

\author{R.~Hartley}
\affiliation{Department of Physics, University of Bath, Bath, BA2 7AY, United Kingdom}

\author{D.~V.~Skryabin}
\affiliation{Department of Physics, University of Bath, Bath, BA2 7AY, United Kingdom}

\author{E.~A.~Cerda-M\'endez}
	\affiliation{Paul-Drude-Institut f\"ur Festk\"orperelektronik, Berlin,
Germany}

\author{K.~Biermann}
	\affiliation{Paul-Drude-Institut f\"ur Festk\"orperelektronik, Berlin,
Germany}

\author{R.~Hey}
	\affiliation{Paul-Drude-Institut f\"ur Festk\"orperelektronik, Berlin,
Germany}

\author{P.~V.~Santos}
	\affiliation{Paul-Drude-Institut f\"ur Festk\"orperelektronik, Berlin,
Germany}

\date{\today}

\begin{abstract}
Microcavity polaritons  are composite half-light half-matter quasi-particles, which have recently been demonstrated to exhibit rich physical properties, such as non-equilibrium Bose-Einstein condensation, parametric scattering and superfluidity. At the same time, polaritons have some important advantages over photons for information processing applications, since their excitonic component leads to weaker diffraction and stronger inter-particle interactions, implying, respectively, tighter localization and lower powers for nonlinear functionality. Here we present the first experimental observations of bright polariton solitons in a strongly coupled semiconductor microcavity.  The polariton solitons  are shown to be non-diffracting high density wavepackets, that are strongly localised in real space with a corresponding broad spectrum in momentum space. Unlike solitons known in other matter-wave systems such as Bose condensed ultracold atomic gases, they are non-equilibrium and rely on a balance between losses and external pumping. Microcavity polariton solitons are excited on picosecond timescales, and  thus have significant benefits for ultrafast switching and transfer of information over their light only counterparts, semiconductor cavity lasers (VCSELs), which have only nanosecond response time.
\end{abstract}

\maketitle


Non-spreading localised wave-packets or solitons occur in a variety of non-linear classical and quantum photonic and matter-wave situations, where they reflect the distinct physical properties of the system under consideration. Light only solitons were first discovered \cite{mol} and most actively researched in optical fibers \cite{agr,rmp}. Fibers  provide a very good  experimental realisation of the one dimensional (1D) Nonlinear Schrodinger (NLS) equation - a paradigm model in soliton physics, which has simple soliton solutions shaped by  opposing dispersive spreading and nonlinearity induced self-phase modulation of optical pulses. Subsequently, the concept of optical solitons has been extended into the spatial domain, where diffraction of propagating beams can be compensated by nonlinear changes  in the material refractive index \cite{agr,sig}. Formally, anomalous group velocity dispersion is equivalent to diffraction: both require positive (self-focusing) Kerr  nonlinearity to form either temporal or spatial bright solitons \cite{agr}.

Recently, matter wave solitons have been demonstrated in Bose condensed atomic gases, where the dispersive spreading of the localized wavepackets induced by the kinetic energy of positive mass atoms is balanced by the attractive interatomic interaction \cite{becsol1,becsol2}.  Interactions between the atoms used in condensation experiments are often repulsive. In order to create bright solitons in these cases, the effective atomic mass must be made negative, as demonstrated using optical lattices \cite{markus}. The same effect can be achieved for light in photonic crystals \cite{moti}. Strong exciton photon coupling in microcavities results in an unusual and advantageous lower polariton branch dispersion exhibiting regions of either positive or negative effective mass \cite{book,pol} depending on the values of transverse momentum \cite{book,pol}. The transition from positive to negative mass is associated with the  point of inflection of the energy-momentum diagram, shown in  Fig. \ref{Fig1}(a). The negative mass of polaritons coupled with repulsive  polariton-polariton interactions favour the formation of the bright solitons which we realise in the present work.

Recent discoveries made with  microcavity polaritons include condensation \cite{Kasprzak2006,Balili2007}, vortices \cite{Lagoudakis2008,Lagoudakis2009}, superfluidity \cite{AmoNP2009}  and follow a pathway similar to the one which led to the observation of coherent matter wave solitons \cite{becsol1,becsol2}. The observation of low threshold  bistability,  polarization multistability \cite{gip,AmoNPhot2010,Sarkar2010,Paraiso} and parametric scattering  \cite{Savvidis} of polaritons  have further prepared the necessary foundations for the realisation of half-light half-matter solitons. In contrast to atomic systems, polariton microcavities operate under non-equilibrium conditions. Atomic systems are typically described by conservative Hamiltonian models, such as NLS or Gross-Pitaevskii equations,  while the polariton system is intrinsically non-Hamiltonian. In the photonic context, this system is often referred to as dissipative, which implies the importance of losses, but also implicitly assumes the presence of an external energy supply \cite{nail,amp}. Theoretically,  conservative half-light half-matter solitons have been predicted in Ref. \cite{saf}, while dissipative bright polariton solitons in microcavities have been numerically studied in Refs. \cite{prl11,prl22}.

\section{Observation of bright polariton solitons}

Here we report  bright solitons in a GaAs semiconductor microcavity (see Methods section for sample and experimental details). The  point of inflection of the polariton dispersion is found at the in-plane polariton momentum $k\simeq2\mu$m$^{-1}$ (see Fig. 1(a)), which corresponds to a group velocity  of noninteracting polaritons  $\partial\epsilon/\partial k\simeq1.8\mu$m/ps, where $\epsilon$ is the polariton energy. We conduct our experiments with polaritons having  momenta above the point of inflection, where the polariton mass  is negative, $M={\hbar^2(\partial^2\epsilon/\partial k^2)^{-1}}\simeq -11.2\cdot10^{-35}$kg ($-1.25\cdot10^{-4}m_e$) at $k\simeq2.4\mu$m$^{-1}$.  As for the lossy and pumped Gross-Pitaevskii equations, see, e.g. Ref.\cite{Skryabin:SolInt}, reasonable estimates of the nonequilibrium soliton parameters can be obtained from the conservative limit of the system.  We can estimate the polariton soliton width for a given polariton density, $N$ using the well-known expression for the healing length of a quantum fluid $w=2\hbar(\sqrt{2Mg N})^{-1}$. This is obtained by equating the characteristic kinetic energy $K$ of the dispersive spreading of the wavepacket with width $w$ ($K=\hbar^2/(2Mw^2)$) to the potential energy $U$ of the polariton-polariton repulsion ($U=g N$, where $g\simeq10$ $\mu$eV$ \mu$m$^{2}$ is the two-body interaction coefficient). For the typical soliton potential energies realised in our experiment $U\simeq0.3$ meV (corresponding to $N\simeq30\mu$m$^{-2}$ ) from which we deduce $w\simeq2\mu m$. The short free polariton lifetime ($\simeq5$ ps) means that solitons emerging from this balance will traverse distances of $\simeq 10\mu$m, before they dissipate. In order to sustain these solitons for longer we need to provide a continuous supply of energy. To sustain these solitons,  energy is pumped into the microcavity using a CW pump beam, focused to a $70 \mu m$ (FWHM) Gaussian spot (Fig. \ref{Fig1}(b)).

We highlight the important properties of dissipative solitons which justify the choice of our experimental parameters to realise soliton production. The polariton density varies strongly across the bright soliton profile reaching its maximum at the center, while the pump beam profile tends to hold the density at a quasi-constant level across the much larger pump spot \cite{prl11,prl22}. Furthermore, due to localisation in real space the soliton profile in momentum space is broad.  As a result, solitons can only be expected under conditions when the pump state is unstable with respect to spatially inhomogeneous perturbations at momenta different from that of the pump \cite{prl11,prl22}. If the microcavity is driven by a pump beam slightly blueshifted with respect to the unperturbed LP branch, the  pump polariton field exhibits bistability as a function of the  pump beam power and angle (Fig. \ref{Fig1}(c), Fig. \ref{Fig4}).  The polariton bistability is usually accompanied by parametric (modulational) instability, which was extensively studied in microcavities \cite{gip,car,kr}. This instability is a particular case of  polaritonic four-wave mixing. Bright solitons are excited on top of the stable background of the lower branch of the bistability loop (Fig. \ref{Fig1}(c)) and can be qualitatively interpreted as locally excited  islands of the modulationally unstable upper branch solution \cite{prl11,prl22}.  Broadband four-wave mixing of polaritons, expanding well beyond the momenta intervals with the parametric amplification, enables coherent scattering from the locally perturbed pump to the   continuum of momenta forming the soliton (Fig. \ref{Fig1}(a)). Simultaneously, the soliton formation requires the transverse momentum of the pump $k_p$, i.e. the incident angle of the laser beam,  to be such that the effective polariton mass is negative, ensuring self-focusing of the repulsively interacting polaritons  \cite{prl11}. This mechanism can lead to self-localisation only along the direction of the pump momentum (1D solitons) \cite{prl11}. Simultaneous localisation in two dimensions involves more subtle physics and the corresponding solitons exist only in a narrow range of pump intensities \cite{prl22}.

Fig. \ref{Fig1}(c) shows the bistable dependence of the polariton emission intensity collected from our device at nearly zero transverse momenta  (direction normal to the cavity plane) as a function of the pump momentum ($k_p$) at the fixed energy of the pump $1.5363$eV, which is $\simeq0.3$meV above the unperturbed LP branch at $k_p\simeq2.37 \mu$m$^{-1}$. With increasing $k_p$ beyond the  bistable interval the intracavity polariton field increases abruptly over the excitation pump spot. This transition is also accompanied by the strong parametric generation of polaritons  into the state with $k\simeq 0$ \cite{Savvidis,gip}. According to the above discussion and the predictions of \cite{prl11,prl22}, bright polariton solitons should exist within the bistability interval (Fig. \ref{Fig1}(c)).

\begin{figure}[h]
\centering
\includegraphics{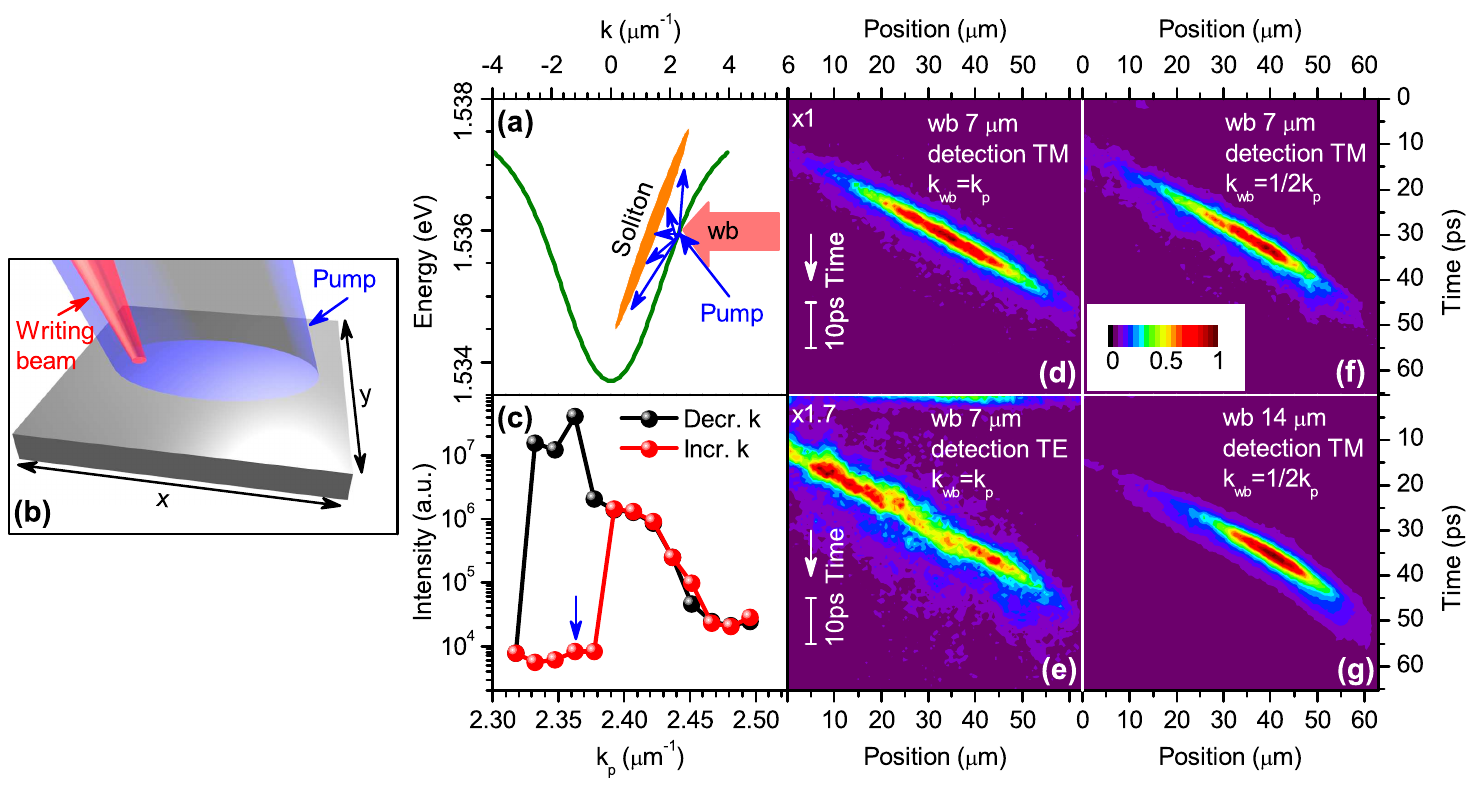}
\caption{(a) Dispersion (energy-momentum) diagram of the lower branch polaritons and schematic representation  of the soliton spectrum and the excitation scheme. (b) Schematic of soliton excitation in the microcavity structure. The CW pump with incident in-plane momentum parallel to the X direction is focused into a large spot. The pulsed writing beam, also incident along X, focused into a small spot triggers soliton formation. (c) Bistable polariton density as a function of the pump momentum; The arrow indicates the state of the system created by the pump before incidence of the writing beam. (d-g) Streak camera measurements of the soliton trajectories along the X direction excited under different conditions. (d) and (e) show the components of the soliton in TE and TM polarizations, respectively. This soliton is excited with a $7\mu$m wide writing beam with in-plane momentum the same as that of the pump. (f) is the same as (d), but the writing beam has half the momentum of the pump. (g) is the same as (f), but the writing beam width is $14\mu$m.}
\label{Fig1}
\end{figure}

We excite bright polariton solitons using a ps writing beam (wb) focused into a spot (Fig. \ref{Fig1}(b)) of diameter in the range $7\mu$m to $15\mu$m,  which is small compared to the diameter of the pump beam of $70\mu$m.  The writing beam is TE and the pump beam is TM polarized. The experimental arrangement of the pump and writing beams is shown schematically in Fig. 1(b), with both beams incident along the X-direction. The unperturbed polariton density (before the application of the writing beam) was in the state corresponding to the lower branch of the bistability loop (see arrow in Fig. \ref{Fig1}(c)) with no indication  of the parametric generation of polaritons with momenta different from the pump. The bistability domain was scanned by changing the pump momentum,  where it was found that the optimal conditions for clear soliton observations exist close to the right boundary of the bistability interval
shown in Fig. \ref{Fig1}(c). The solitonic emission was collected along a $2\mu$m stripe of the streak camera image (see Methods) parallel to the direction of incidence of the pump. The light was collected inside the finite interval of angles corresponding to momenta $0<k<k_p$, thereby avoiding collection of the reflected pump beam, which otherwise leads to detector saturation. Figs. \ref{Fig1}(d) and (e) show spatio-temporal traces of the intensities in TE and TM polarisations, for typical non-diffracting and non-decaying propagating wavepackets excited with the writing beam. Here the $7\mu$m writing beam arrives at position $X=-20\mu$m at time, $t=0$ where the transverse momentum is the same as that of the pump $k_{wb}=k_p$.

We now present the experimental results, supported by theory in the next section, which prove that the wavepackets arise from bright polariton solitons propagating across the excitation spot. Firstly, we demonstrate that the velocity of the soliton is independent of $k_{wb}$. The role of the writing beam is to create a local perturbation of the pump state, which in turn results in soliton formation due to scattering from the pump state (Fig. \ref{Fig1}(b)). It has been shown numerically for linearly polarised polariton solitons that their velocity is close to the group velocity of the polaritons at the pump momentum \cite{prl11}. To test this prediction, we changed the momentum of the writing beam to half of the pump momentum, $k_{wb}=k_p/2$ (Figs. 1(f,g). We observe a soliton velocity in Fig. \ref{Fig1}(f) of $\simeq 1.68\mu$m/ps, the same as that in Figs. \ref{Fig1}(d) and (e), where $k_{wb} = k_p$. The independence of the polariton soliton velocity on $k_{wb}$ is in sharp contrast to that expected for conservative solitons, where the soliton velocity is solely determined by the momentum of the  excitation pulse \cite{agr}.

\begin{figure}[t]
\centerline{\includegraphics[width=11cm]{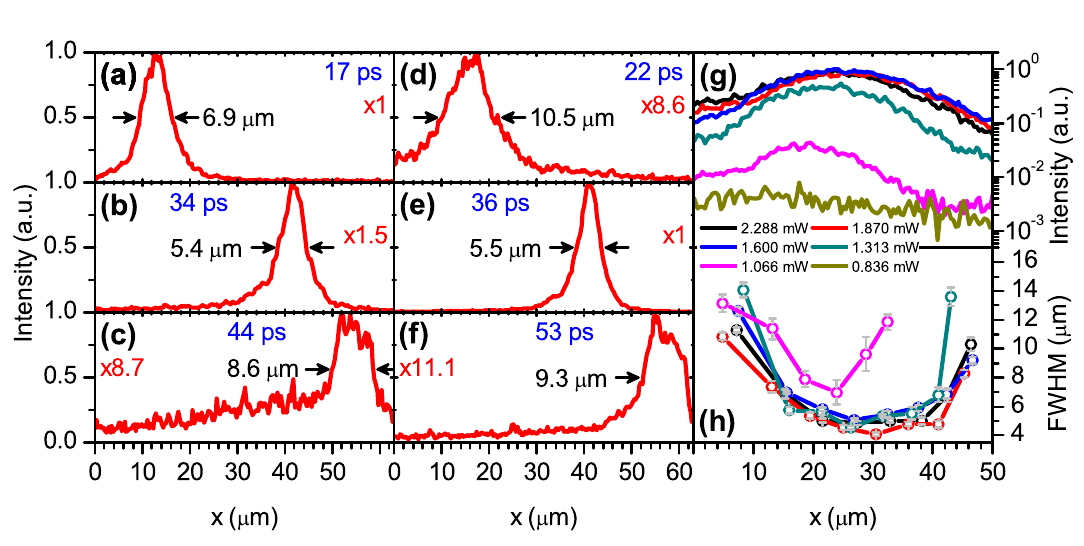}}
\caption{(a)-(c) Experimentally measured spatial intensity profiles of a soliton created by the $7\mu$m writing beam at different times, showing excitation and decay of the soliton; (d)-(f) Intensity profiles of a soliton created by the $15\mu$m writing beam; (g,h) Dynamics of the peak intensity and width of the wavepackets excited by the $7\mu$m writing beams with different powers as they propagate across the cavity. The writing powers where the above parameters remain quasi-constant in the interval of $20$ to $40\mu$m correspond to the formation of solitons.}
\label{Fig2}
\end{figure}

Secondly, we show that the size of the soliton is determined by the pump and cavity parameters and is independent of the size of the writing beam, as is true for other types of dissipative solitons \cite{nail,amp}. The size of the excited wavepacket ($w$) in the soliton regime is expected to be fixed by the potential energy $U$ of the solitons, where $U$ is of the order of the pump energy detuning with respect to the energy of the unperturbed lower branch polaritons ($w\sim 1/\sqrt{U}$ as discussed earlier). This is illustrated in Figs. \ref{Fig2}(a-f), which show the profiles of polariton wavepackets along their propagation direction (X) at different times and positions for writing beam sizes of $\simeq7 \mu$m  (Fig. \ref{Fig2}(a-c)) and  $\simeq15 \mu$m (Fig. \ref{Fig2}(d-f)).  In the initial stage of the soliton excitation, these writing beams produce polariton wavepackets of very different widths (see Figs. \ref{Fig2}(a) and (d)), which then quickly evolve into solitons of the same size $\simeq5 \mu$m as shown in Figs. \ref{Fig2}(b) and (e), as expected for soliton formation.

Thirdly, we  demonstrate the unambiguous  non-spreading and non-decaying features  of the soliton wave packets. Figures \ref{Fig2}(g) and (h) show the dependences of the intensity and width of the excited wavepackets, respectively, versus their position as a function of the writing beam power, $P_{wb}$. Within the range of $P_{wb}$ from 1.6 mW to 2.3 mW, the soliton intensity and width are nearly constant. Furthermore, the width versus position dependence (Fig. \ref{Fig2}(h)) exhibits a plateau in the interval from $\simeq 20$ to $\simeq 40\mu$m. These signatures of polariton soliton formation are further discussed in the modeling section. At the edge of the pump spot, the pump intensity is insufficient to maintain the system within the bistability region and thus to sustain solitons. This is the main factor determining  the practical extent of the soliton trajectory, and leads to spreading and dissipation of the excited wavepackets  at times greater than 40-45 ps when the edge of the pump spot is approached, as shown in Figs. \ref{Fig1}(d-g) and Fig. \ref{Fig2}(c),(f).

We note that the intensity of the writing beam must be high enough to enable soliton switching. (Fig. \ref{Fig2}(h)). As one reduces the pump intensity below the threshold of $1.3$mW (Fig. \ref{Fig2}(h)) for soliton switching, we observe an abrupt decrease in the intensity and an increase in width of the wavepacket.  Further reducing the pump intensity $P_{wb} < 1mW$, only quickly decaying non-solitonic wavepackets are excited. For pump  momenta that are below the point of inflection, the writing beam triggers switching of the whole pump spot \cite{AmoNPhot2010} and no soliton formation is observed.

\begin{figure}[t]
\centering
\includegraphics{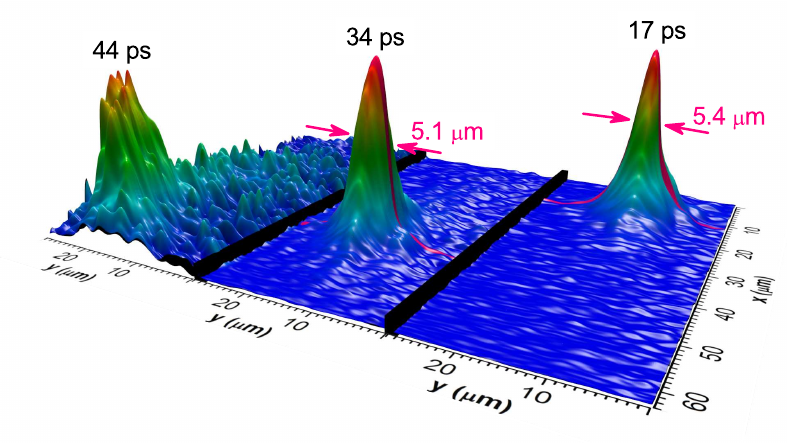}
\caption{Two dimensional streak camera measurements of a soliton traveling across the microcavity plane. Experimental conditions are the same as in Figs. \ref{Fig2}(a-c). Cross sections taken along the Y direction indicates a soliton full width at half maximum of $5\mu$m.}
\label{Fig3D}
\end{figure}

Figure \ref{Fig3D} shows two dimensional images of solitons. The soliton size along the Y coordinate perpendicular to the propagation direction (see Fig. \ref{Fig1}(b)) at $30-40$ps is  $\simeq 5\mu$m, narrower than the initial beam size. This  indicates suppression of diffraction and localization in the direction  where polaritons have a positive effective mass. Whereas localisation in the X direction described above arises from the interplay between the negative effective mass and repulsive interactions, the origin of localisation along  Y  is different. It can be interpreted in terms of the  phase dependent parametric nonlinearity and the interaction of propagating fronts, see Ref. \cite{prl22}.

We note that our observations have a very different physical origin to the triggered optical parametric oscillator observations of Ref. \cite{amo_topo}. In that case, the system is already in a high density phase before a writing beam is applied \cite{amo_private} and furthermore the system  was pumped at low $k_{p}$ in the region of positive effective mass, where soliton generation cannot occur. As a result, propagating  excitations of the polaritonic condensate were studied, as opposed to the bright solitons  reported by us (see Supplementary Information for further discussion).

\section{Numerical modeling of soliton formation}

In order to provide further insight into our experimental observations, we have performed a series of numerical simulations. This is important since the previous theoretical studies  of polariton solitons \cite{prl11,prl22} are restricted to the linearly polarized case, while in the present setting, with orthogonally polarised pump and writing beams, coupling between the two polarisations is important and leads to the formation of elliptically polarised solitons. The model we have used includes equations for the TE (polarisation parallel to the cavity plane) and TM polarised optical modes and for the respective excitonic fields (see Methods for details).

\begin{figure}[t]
\centerline{\includegraphics{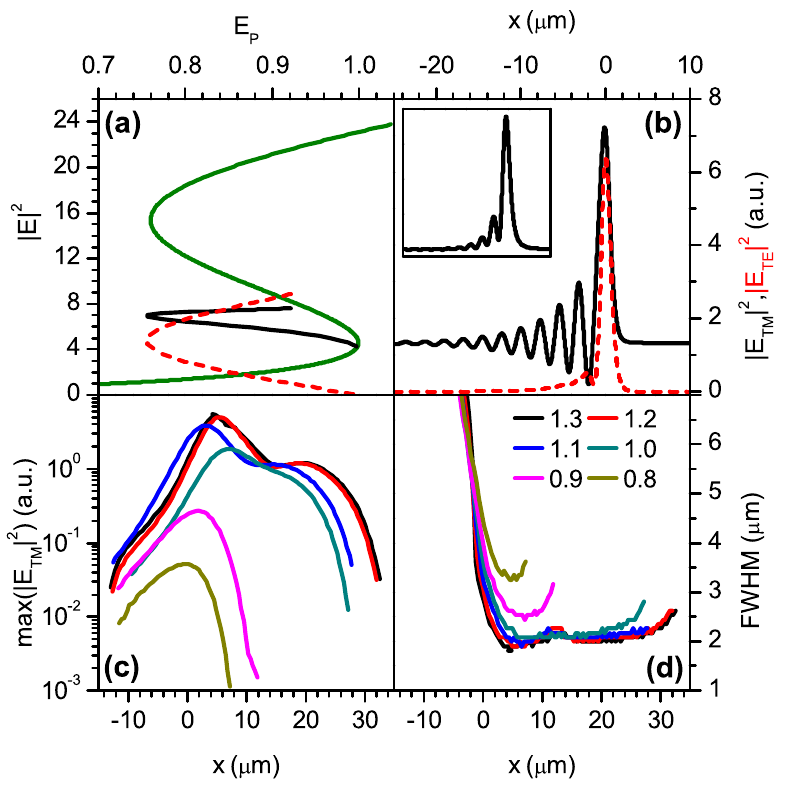}}
\caption{Numerically computed ideal solitons (a,b) and dynamics of the soliton excitation (c,d). (a) The full green line shows the bistable spatially homogeneous polariton state. The full black line and the dashed red line show the amplitudes of the TE and TM components of the elliptically polarized soliton, respectively. (b) Spatial profiles of the TE (full black) and TM (dashed red) soliton components. The inset shows the soliton profile when the light with $k\ge k_p$ has been filtered out. (c) and (d) Dynamics of the soliton width and peak intensity as functions of position across the microcavity. The writing beam is $7\mu$m wide and its duration is $1$ps. Different line colors correspond to the different intensities of the writing beam indicated in the inset in (d). One unit of all the plotted dimensionless field amplitudes corresponds to the pump beam amplitude $E_p$ at the right end  of the lower branch of the bistability loop, see (a).}
\label{Fig4}
\end{figure}

Transforming from the laboratory frame to the frame moving with an unknown velocity, which is consistently determined together with the soliton profile \cite{prl11}, we have found branches of elliptically polarised solitons, see Figs. \ref{Fig4}(a,b). These solutions retain their spatial profile and all other characteristics for indefinitely long times. The soliton branches with positive slope in Fig. \ref{Fig4}(a) represent  propagating solutions, which are stable with respect to perturbations. The negatively sloped branches correspond to unstable solitons. These ideal solitons have been calculated on top of the spatially homogeneous pump and move with velocities $\simeq 1.7\mu$m/ps, in good agreement with experiment.

Another series of numerical simulations has been conducted to demonstrate the  excitation of solitons with $1$ps pulses on top of the gaussian pump beam, see Figs. \ref{Fig4}(c,d). Fig. \ref{Fig4}(c) (cf. Fig. \ref{Fig2}(g)) and Fig. \ref{Fig4}(d) (cf. Fig \ref{Fig2}(h)) show changes of the peak intensity and of the width of wavepackets traveling across the microcavity for several intensities of the writing beam. The saturation of both parameters with increasing intensity of the writing beam (Fig. \ref{Fig4}(c)) as well as the quasi-constant soliton width in the region from 5 $\mu$m to $30$ $\mu$m of the propagation length (Fig. \ref{Fig4}(d)) demonstrate the transition to the soliton regime, and is fully consistent with the experimental observations in Figs. \ref{Fig2}(g-h). The observed saturation behaviour indicates that as for other dissipative solitons \cite{nail,amp} the soliton width in our case is fixed by the pump and cavity parameters\cite{prl11,prl22} and not by the writing beam powers. This is in strong contrast to conservative solitons, which for the given system parameters can exist with an arbitrary width determined by the density $N$ initially induced by the writing beam.

\section{Soliton observation in energy-momentum space}

\begin{figure}[t]
\centering
\includegraphics{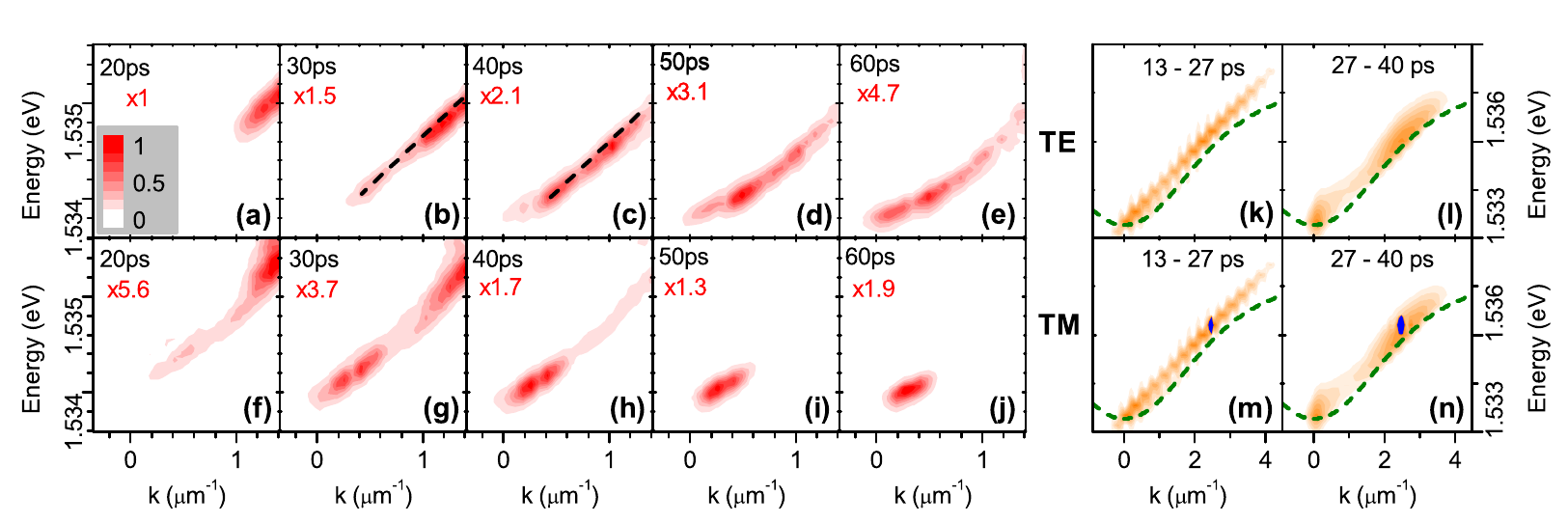}
\caption{
Images of polariton emission in energy-momentum space taken at different times after the application of the writing beam in TM (bottom panel) and TE (top panel) polarisations. (a-j) Experimental data. Dashed line is a linear fit to the experimental measurements. The energy resolution of 0.1 meV limits the time resolution to $\approx10$ ps. (k-n) Numerical simulations of polariton emission in time intervals of $13-27$ ps (k,m) and $27-40$ ps (l,n) after the application of the writing beam. Note, the span in the momentum space is wider than in the experimental figure. Dashed lines show the dispersion of the noninteracting LP branch.
}
\label{Fig5}
\end{figure}

The soliton formation can also be viewed as scattering of pump polaritons into a continuum spanning a broad range of transverse momenta from $0$ to $2k_p$ (see Fig. \ref{Fig1}(a)) \cite{prl11}. Our measurements of the energy-momentum ($E(k)$) profile of the solitonic wavepackets over time, as shown in Fig. \ref{Fig5}, confirm this point. Initially, for $t<20$ps, the polariton emission is concentrated mostly close to $k_p$, whereas at later times ($30-40$ ps) the emission is distributed over a broad range of momenta a characteristic feature of soliton formation \cite{prl11}. The soliton spectra are expected to form a straight line  tangential  to the dispersion of nonsolitonic radiation \cite{rmp,prl11}. Indeed, the  measured dispersion at $30-40$ps can be approximated by a straight line (dashed lines in Fig. \ref{Fig5}(b,c)) in agreement with our numerical modeling, see Fig. \ref{Fig5} (k,m). We note however that soliton emission forms at $k$-vectors $k\geq0.5$ $\mu$m$^{-1}$ and spans over the point of inflection, where the dispersion of non-interacting polaritons is also close to linear. Nevertheless, the difference between the solitonic dispersion observed in experiment at around $30-40$ps (in modeling at $13-27$ ps) and the dispersion of the nonsolitonic polaritons observed at later times $\geq50$ps (in modeling at $27-40$ps) is clear from the data shown in Fig. \ref{Fig5}. The soliton velocity, $\frac{1}{\hbar}\frac{\partial\epsilon}{\partial k}$, deduced from the dashed lines in Figs. \ref{Fig5}(b, c) is  $\simeq 1.6\mu$m/ps, consistent with the measurements in Fig. \ref{Fig1} and the numerical modeling. The difference between the experimental TE and TM emission in momentum space may arise from anisotropy of polariton scattering dynamics (see for example \cite{DNK:Rot}), which is not taken into account in our modeling.

\section{Discussion and conclusions}

The experimental observations of bright polariton solitons   reported above open opportunities for the exploration  of their potential applications in ultrafast information processing, since their picosecond response time is three orders of magnitude faster than that observed for the pure light cavity solitons in VCSELs \cite{nature,apl,amp}. Furthermore polaritonic nonlinearities are  2-3 orders of magnitude larger than nonlinearities in VCSELs \cite{nature,apl,amp}. The measured transverse dimensions of polaritonic solitons are $\simeq 5\mu$m (resolution limited), whereas the numerical model using the  experimental parameters  predicts  $\simeq 2\mu$m, several times less than the $10\mu$m width typical for  VCSEL solitons. Polariton solitons can potentially be  used as more natural information bits  than the propagating domain walls in the recently proposed integrated polariton optical circuits and gates (polariton neurons) \cite{LiewNeurons}. The number of polaritons in the solitons observed is of the order of hundreds, which puts them into the category of mesoscopic structures. Realisation of spatially modulated 1D microcavity structures, with reduced numbers of particles may create  conditions for observation of quantum solitons, and nonlinear functionality relying on few polariton quanta, potentially allowing designs of quantum polariton-soliton  circuits. We note finally that unlike the well-studied polariton condensates which correspond to macroscopic occupation of a single state in momentum space, the highly occupied polariton soliton is strongly localised in real space with a broad spread in energy and momentum.

In parallel with our work, Amo \textit{et. al.} have very recently reported observations of dark 1D polariton solitons \cite{amo}. These solitons are formed by polaritons with positive mass in collision with obstacles. A most important physical difference with the bright polariton solitons reported here arises from the fact that in the experiment of Ref. \cite{amo}, the solitons are observed outside the pump spot, necessary to avoid locking of the dark soliton phase to the pump phase in their one-beam experiments. Such an arrangement selects conservative solitons propagating in the presence of the unbalanced absorption.

\section{Methods}

We use a GaAs-based device with six $15$nm thick quantum wells (QWs) grown by molecular beam epitaxy. The experiments were performed at a temperature of $5$K. Propagation of the wave packets in real time was recorded using a Hamamatsu Streak Camera with a time resolution of $2$ps. An aspheric lens with a focal length of 4 cm and numerical aperture $NA=0.44$, provided optical resolution of about 4-5 $\mu$m for both soliton excitation and imaging. Two-dimensional images of solitons at different times were reconstructed from an array of 1D images versus position X  such as in Fig. 1 obtained at different positions Y on the pump spot. Solitons were driven by a TM polarised CW pump beam and triggered by a pulsed TE polarised writing beam with duration of 5 ps. The pump and writing beams were directed to the microcavity through the same optical path using a polarising beam splitter. The orthogonal polarisations of the beams were chosen to ensure transmission of maximum available powers from each of the lasers to the sample.

Numerical studies were performed using mean-field equations describing the evolution of slowly varying amplitudes of the TM and TE cavity modes and of the corresponding excitonic fields $\psi_{TM,TE}$:
{\small
\begin{eqnarray}
&& \partial_t E_{TM} - i\frac{\hbar}{2m_c}\left(\partial^2_x+\partial^2_y\right)E_{TM}
+\left(\gamma_c-i\delta_c\right)E_{TM}=i\Omega_R\psi_{TM}+E_p(x,y)e^{i k_p x}\;,
\label{eqEx}\\
&& \partial_t E_{TE} - i\frac{\hbar}{2m_c}\left(\partial^2_x+\partial^2_y\right)E_{TE}
+\left(\gamma_c-i\delta_c\right)E_{TE}=i\Omega_R\psi_{TE}
-iE_{wb}(x,y,t)e^{i\kappa_{wb} x}\;,
\label{eqEy}\\
&& \partial_t \psi_{TM} +\left(\gamma_e-i\delta_e\right)\psi_{TM}+\frac{ig}{4}\left[(1+r)|\psi_{TM}|^2+2|\psi_{TE}|^2\right]\psi_{TM}
-\frac{ig}{4}(1-r)\psi_{TM}^*\psi_{TE}^2=i\Omega_R E_{TM}\;,
\label{eqPsix}\\
&& \partial_t\psi_{TE} +\left(\gamma_e-i\delta_e\right)\psi_{TE}+\frac{ig}{4}\left[(1+r)|\psi_{TE}|^2+2|\psi_{TM}|^2\right]\psi_{TE}
-\frac{ig}{4}(1-r)\psi_{TE}^*\psi_{TM}^2=i\Omega_R E_{TE}.
\label{eqPsiy}
\end{eqnarray}
}
Here  $m_c=0.27\cdot 10^{-34}$kg is the effective cavity photon  mass, $\hbar\Omega_R=4.9867$meV is the Rabi splitting, $\hbar\gamma_c=\hbar\gamma_e=0.2$ meV are the cavity photon and the exciton coherence decay rates, $\delta_e=-1.84$ meV, $\delta_c=-2.34$ meV, $g>0$ is the nonlinear parameter, which can be easily scaled away, $r=-0.05$  parameterizes the nonlinear interaction between the two modes \cite{kavok}. $E_p(x,y)$ is the pump amplitude with the  momentum $k_p$, the corresponding the angle of incidence $\theta=\arcsin[\kappa\lambda_p/(2\pi)]$ and $E_{wb}$ is the writing beam amplitude.

For the case of the homogeneous pump $E_p(x,y)=const$ and $E_{wb}(x,y,t)\equiv 0$ the soliton solutions are sought in the form $E_{TM,TE}=A_{TM,TE}(x-vt)e^{i k_p x}$, $\psi_{TM,TE}=Q_{TM,TE}(x-vt)e^{i k_p x}$. The soliton profiles and the unknown velocity $v$ are found self-consistently using Newton-Raphson iterations. In our simulations of the soliton excitation the system of Eqs.~(\ref{eqEx})-(\ref{eqPsiy}) has been solved directly using the split-step method.

\section{Acknowledgements}
The Sheffield group thanks EPSRC (EP/G001642), the FP7 ITN Clermont 4 and the Royal Society for support of this work, and A.~Amo for a helpful discussion.

\section{Contributions}
All authors prepared the manuscript and analysed experimental and numerical data;
M.S. and D.N.K. conducted the experimental measurements;
A.V.G., R.H., D.V.S. conducted the theoretical and numerical work;
D.V.S. proposed the concept;
K.B. and R.H. fabricated the microcavity.

\section{Competing financial interests}
The authors declare no competing financial interests.

\section{Corresponding authors}

D.~N.~Krizhanovskii and D.~V.~Skryabin

\renewcommand{\figurename}{FIG. S}
\setcounter{figure}{0}
\newpage

\section{Supplementary Materials: Propagating Polariton Wavepacket in Triggered Optical Parametric Oscillator Configuration}

As was discussed in the main text, optical bistability of the pump field together with the negative polariton effective mass at $k_{p}$ are necessary conditions for formation of a stable solitonic wave. In this case the soliton results from switching in the bistable region ($k_{p}\simeq2.38\mu m^{-1}$) from the lower to the upper state locally in a region of a few microns. By contrast, if the whole pump spot is switched on, i.e the system is in the upper state in the bistability region or the k-vector of the pump is above the upper bistability threshold then a soliton can not be formed. This is demonstrated in the following experiment.

Firstly, we set the system to the upper state (Fig. 1(c) of the main text), where a condensed 'signal' state is formed at $k=0$, with long range spatial coherence over $30\mu m$, due to polariton-polariton parametric scattering from the pump state into 'signal' and 'idler' states at $k_{signal}=0$ and $k_{idler}=2k_{p}$. This corresponds to the polariton optical parametric oscillator (OPO), which was studied in Ref.~\cite{Stevenson_OPO}.

Secondly, we introduce a pulsed writing beam (wb) at $k_{wb}\simeq1.2\mu m^{-1}$ focused to a small spot of $\simeq7\mu m$ and observe propagation of the resulting wavepacket. The wavepacket is amplified due to polariton-polariton scattering from the switched on pump state and hence propagates macroscopic distances, superimposed on the signal condensate at $k=0$. Such a process corresponds to the so-called triggered OPO (TOPO), firstly reported by Amo et. al. in Ref.~\cite{amo_topo}.

The TOPO propagating wavepacket exhibits very different physical properties to the soliton. Fig. S1(a) shows the TOPO intensity as a function of time and position $X$ (see Fig.~1(b) of the main text) for $k_{p}\simeq 2.38 \mu m^{-1}$. It is seen that the TOPO wavepacket propagates over long distances of up to $50\mu m$. However in contrast to the soliton propagation, it broadens significantly from $7 \mu m$ up to $15-20\mu m$ within $\simeq40 ps$ time and propagates at a speed 1.3 times slower than the observed soliton. Moreover the TOPO wavepacket intensity decays with time, again in marked contrast to soliton behaviour of Fig. 2(g).

The TOPO wavepacket created by the writing beam can be considered as a condensate excitation, which has the properties of as a diffusive Goldstone mode \cite{Wouters:GoldstoneMode}. There is competition between polariton-polariton scattering from the pump to the signal state at $k=0$ and to the localised TOPO wavepacket at $k_{wb}$. Since the pump state is switched on over the whole excitation spot there are no excitations in the pump state itself. The TOPO wavepacket propagation speed is determined only by the writing beam k-vector $k_{wb}\simeq1.2\mu m^{-1}$ and not by that of the pump as in the case of soliton. The TOPO wavepacket expands since the polariton dispersion at $k_{wb}\simeq1.2\mu m^{-1}$ is described by a positive effective mass.

The time dependence of the TOPO wavepacket FWHM is expected to be strongly dependent on the initial size of the writing beam. It can be easily shown from solution of a time dependent Schrodinger equation that the FWHM of a noninteracting wavepacket with initial FWHM $\Delta$ with a dispersion described by positive effective mass $m_{eff}$ is given by $2\sqrt{\Delta^2/4+\frac{\hbar^2}{m_{eff}^2\Delta^2}4t^2}$. It is seen that the size of the wavepacket will increase with time more slowly for initially larger $\Delta$. This is consistent with the above observation (wavepacket with initial size $7\mu m$ is expected to spread up to $\simeq20\mu m$ within 50 ps) and those in Ref.~\cite{amo_topo}, where the TOPO wavepacket of a size 15-20 $\mu m$ along the propagation direction is initially created and its size changes very little within 40-50 ps. Indeed, using the above formula we can estimate that it should increase only by 2 microns for $m_{eff}\simeq10^{-34}$ kg.

\begin{figure}
\centering
\includegraphics{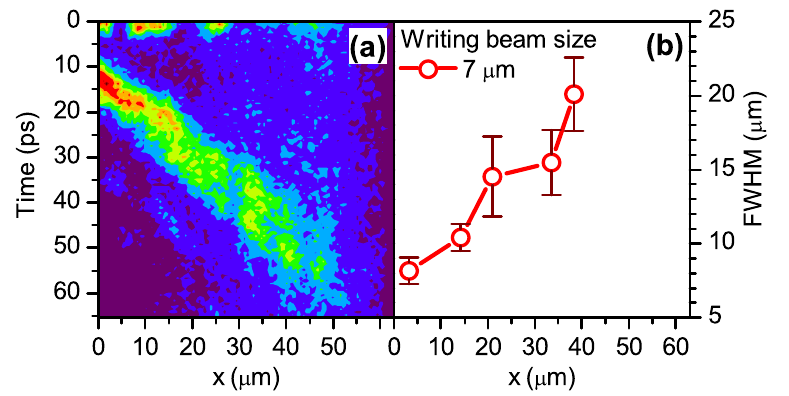}
\caption{(a)- streak camera trace of propagating TOPO wavepacket as a function time and position recorded for $k_{p}\simeq 2.38\mu m^{-1}$ and writing beam size $7\mu m$. (b) - size (FWHM) of TOPO wavepacket recorded in panel (a) as a function of position.}
\label{Fig1}
\end{figure}

\end{document}